\documentclass[preprint]{vgtc}               

\ifpdf
  \pdfoutput=1\relax                   
  \usepackage{graphicx}                
  \DeclareGraphicsExtensions{.pdf,.png,.jpg,.jpeg} 
\else
  \usepackage{graphicx}                
  \DeclareGraphicsExtensions{.eps}     
\fi%

\graphicspath{{figures/}{pictures/}{images/}{./}} 

\usepackage{microtype}                 
\PassOptionsToPackage{warn}{textcomp}  
\usepackage{textcomp}                  
\usepackage{mathptmx}                  
\usepackage{times}                     
\usepackage{cite}                      
\usepackage{tabu}                      
\usepackage{booktabs}                  

\onlineid{0}

\vgtccategory{Research}

\vgtcinsertpkg

\preprinttext{Author’s version. Final record to appear in 2023 IEEE International Symposium on Mixed and Augmented Reality Adjunct (ISMAR-Adjunct).}

\title{Feel the Breeze: Promoting Relaxation in Virtual Reality using Mid-Air Haptics}

\author{Naga Sai Surya Vamsy Malladi\thanks{e-mail: naga.malladi@uni-bayreuth.de}\\ 
     \scriptsize University of Bayreuth %
\and Viktorija Paneva\thanks{e-mail: vpaneva@acm.org}\\ %
        \scriptsize University of Bayreuth %
\and J{\"o}rg M{\"u}ller\thanks{e-mail: joerg.mueller@uni-bayreuth.de}\\ %
     \parbox{1.4in}{\scriptsize \centering University of Bayreuth \\ 
     }}

\abstract{
Mid-air haptic interfaces employ focused ultrasound waves to generate touchless haptic sensations on the skin. 
Prior studies have demonstrated the potential positive impact of mid-air haptic feedback on virtual experiences, enhancing aspects such as enjoyment, immersion, and sense of agency. 
As a highly immersive environment, Virtual Reality (VR) is being explored as a tool for stress management and relaxation in current research. 
However, the impact of incorporating mid-air haptic stimuli into relaxing experiences in VR has not been studied thus far. 
In this paper, for the first time, we design a mid-air haptic stimulation that is congruent with a relaxing scene in VR, and conduct a user study investigating the effectiveness of this experience.
Our user study encompasses three different conditions: a control group with no relaxation intervention, a VR-only relaxation experience, and a VR+Haptics relaxation experience that includes the mid-air haptic feedback. 
While we did not find any significant differences between the conditions, a trend suggesting that the VR+Haptics condition might be associated with greater pleasure emerged, requiring further validation with a larger sample size. 
These initial findings set the foundation for future investigations into leveraging multimodal interventions in VR, utilising mid-air haptics to potentially enhance relaxation experiences.
}

\CCScatlist{
  \CCScatTwelve{Human-centered computing}{Human computer interaction (HCI)}{Interaction paradigms}{Virtual reality};
  \CCScatTwelve{Human-centered computing}{Human computer interaction (HCI)}{HCI design and evaluation methods}{User studies}
}

\begin{document}

\firstsection{Introduction}
\maketitle

Stress is the human body's way of responding to any stimulus that could be perceived as a threat. 
Prior research demonstrates that stress can have a negative impact on memory, cognition, and learning~\cite{Pascoe2020, Yaribeygi2017}. 
Notably, an extensive study conducted as part of the World Health Organisation World Mental Health International College Student Initiative, involving over 20,000 students across 9 countries and 24 universities, showed that 93.7\% of the respondents experienced at least some stress in at least one life area~\cite{Karyotaki2020}. 
The study also identified a causal effect of stress on six types of mental health disorders, further underlining the compelling need to investigate strategies for stress management and prevention.
In particular, there is a need for a user-friendly stress management solution that can be easily incorporated into daily life. 
Furthermore, the solution should be easily accessible to nonclinical groups and potentially complement other forms of treatments and interventions.
Due to its immersive nature, VR holds promise as a viable tool for fostering relaxation and mitigating stress.
However, most VR relaxation applications are based only on visual and auditory modalities, overlooking the potential benefits of touch (e.g., in perceiving and communicating emotions~\cite{Grunwald2008}). 

In this paper, we introduce a multimodal relaxation experience that incorporates a serene VR scene with congruent mid-air haptic sensations.
Utilising mid-air haptic technology, tactile sensations are generated directly on the user's skin without physical contact. 
We hypothesise that integrating visual, auditory, and haptic stimuli to create a congruent multimodal experience in VR will result in heightened relaxation and more effective stress reduction. 
We conduct a comprehensive evaluation to investigate the effectiveness of our approach in reducing stress and promoting relaxation, compared to a control condition with no relaxation intervention, and to a purely VR experience with no haptic feedback.
The findings of this study contribute to the growing field of stress management and relaxation in virtual environments, with potential applications in diverse settings, from well-being interventions to therapeutic contexts.

\section{Related Work}

\subsection{VR for Stress Management and Relaxation}
VR with its ability to transport users to fully immersive environments that are vastly different from the real world, has emerged as a powerful tool for applications ranging from exposure-based therapy to gaming and simulation~\cite{Emmelkamp21, Paneva2020LeviSim}. 
Recent studies have leveraged this immersive property to design and investigate different relaxation techniques in VR, for example in combination with meditation~\cite{Rakowski2021} or breathing exercises~\cite{Soyka2016}.
VR relaxing experiences involving natural scenery deployed during work breaks, have shown to be effective forms of stress management and relaxation among office workers~\cite{Thoondee2017}, clinicians~\cite{Adhyaru2022}, and college students~\cite{Rakowski2021}. 
Most of the existing VR relaxation applications in the literature primarily rely on the visual and auditory modalities, with the exception of a few studies that have incorporated olfactory interfaces~\cite{Amores2018}.
To our knowledge, no VR application that incorporates mid-air haptic stimulation has been designed and tested thus far for the purpose of stress reduction and relaxation.

\subsection{Haptics for Stress Management and Relaxation} Touch is an important communication modality that can have therapeutic effects and hence plays a role in many relaxation interfaces~\cite{Im2019}.
A variety of haptic devices have been designed to promote relaxation, including Good Vibes~\cite{Kelling16}, which is a haptic sleeve containing vibro-tactile actuators that use dynamic vibration patterns as a soothing intervention in stressful situations.
Bumatay et al.\cite{Bumatay15} designed a mobile meditation tool and a breathing guide involving a vibrating huggable pillow, while Haynes et al.\cite{Haynes22} recently introduced a huggable haptic interface that uses pneumatics, instead of vibration motors, to more closely simulate slow breathing.
However, wearable or handheld devices often provide a limited variety of haptic sensations. 
In contrast, EmotionAir~\cite{Tsalamlal13} proposes non-contact tactile stimulation using a rotatable air nozzle and investigates how different parameters of the airflow relate to the valence, arousal, and dominance dimension of the affective scale.
Although the air jet haptic display can cover relatively large areas of the body, e.g., the forearm, it has a limited resolution. 
For a more detailed overview on affective haptic interfaces, please see Eid et al.~\cite{Eid16}.

\subsection{Mid-Air Haptic Interfaces}
Mid-air haptic technology offers a touchless and hygienic method for providing haptic feedback to the user, as it operates at a distance without requiring any direct physical contact with the user~\cite{iwamoto2008non, ultrahaptics}. 
This technology employs an array of ultrasonic transducers to emit ultrasonic waves, modulated down to a frequency perceptible by the mechanoreceptors in the human skin~\cite{ultrahaptics}. 
The parameters of the array can be adjusted to generate various haptic patterns, shapes, and textures~\cite{Paneva2020, Hajas20, Freeman17}. 
Ultrasonic mid-air haptic displays offer a high spatial and temporal resolution, operating at a distance, and enabling users to interact with them using only their bare hands, without the need for wearables or attached gadgets.

\subsection{Affective Mid-Air Haptics}
Findings of prior studies indicate that it is possible to convey emotional meaning using ultrasonic mid-air haptics~\cite{Obrist15}.
Seinfeld et al.~\cite{Seinfeld22} found that visuotactile feedback using mid-air haptic stimulation, has a positive effect on the illusion of being affectively touched by a virtual avatar and embodied in a virtual body in VR, compared to a purely visual condition.
An audio-haptic demonstrator was developed to augment emotional short story narratives through mid-air haptics~\cite{Sheremetieva22}.
Tsumoto et al.~\cite{Tsumoto21} investigated the effect of different spatotemporal parameters of the mid-air haptic stimulus on the perceived pleasantness.
Moreover, further studies have revealed that adding mid-air haptic stimulus can enhance participant immersion in art experiences~\cite{Vi17}, increase enjoyment when interacting with digital signage~\cite{Limerick2019}, positively influence emotions when watching short movies~\cite{Ablart2017}, and increase the implicit sense of agency during a virtual button press~\cite{Evangelou21}.

\section{Methods}

\subsection{Experimental Design}
To control for any lasting effects both of the stress inducing and the relaxation intervention, we opted for a between group experimental design. 
We note that this is an accepted practice for relaxation studies in the literature~\cite{Bumatay15, Kelling16, Strassmann19, Rakowski2021}.
Participants were randomly assigned to one of the following three conditions: 1) No relaxation experience (\textit{Control}), 2)  Relaxation experience in VR (\textit{VR-only}), and 3) Relaxation experience in VR with mid-air haptic feedback (\textit{VR+Haptics}).

\subsection{Participants}
A total of 24 participants (13 female, 10 male, and 1 other) with a mean age of 25.75 (SD=2.44) were recruited for the experiment.
Most of the participants were students, while the rest were university employees.
The study 
followed ethical standards as per the Helsinki Declaration. 
All participants received a monetary reimbursement for their participation.
\begin{figure}[t]
  \centering
  \includegraphics[width=0.6\linewidth]{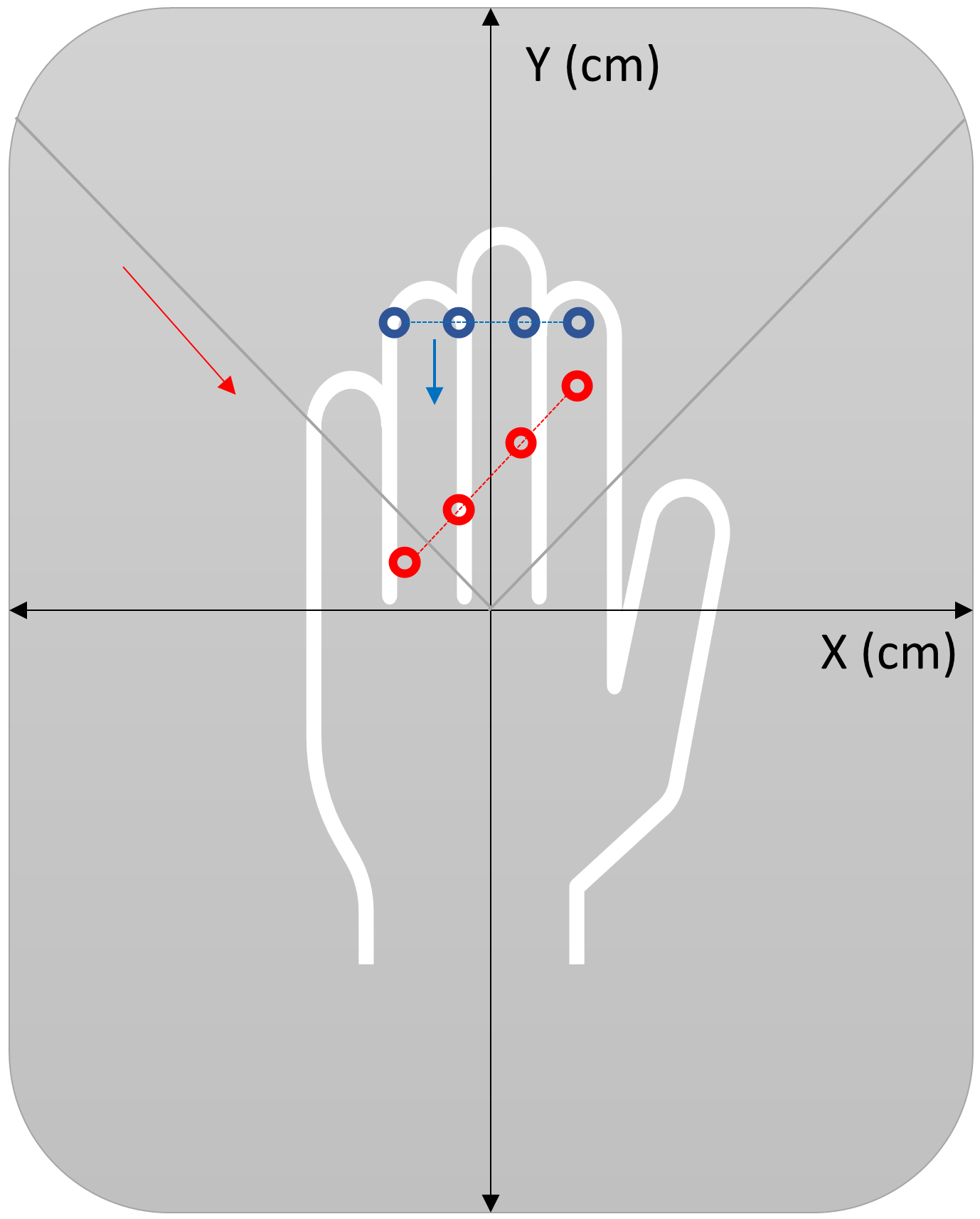}
  \caption{Illustration depicting our mid-air haptic representation of a sea breeze. The red and blue circles are two different examples of possible sets of haptic control points in a single iteration. The grey lines flanking the y-axis represent the permissible range of angles that the direction of motion of the haptic stimulation line can make with the y-axis. This angle is randomly varied every third iteration. The blue arrow indicates the direction of motion of the blue haptic points (oriented at a 0° angle relative to the y-axis), while the red arrow indicates the direction of the red points (oriented at a 45° angle relative to the y-axis). 
  }
  \label{fig:hand}
\end{figure}

\subsection{Experimental Stimuli}

In both the VR-only and VR+Haptics conditions, we utilised a Vive Pro\footnote{www.vive.com} head mounted display. 
The mid-air haptic stimulus in the VR+Haptics condition was generated using an Ultraleap STRATOS Explore development kit\footnote{www.ultraleap.com/haptics}.
The therapeutic effects of natural environments, particularly those associated with water bodies, is well-documented in the literature~\cite{Anderson2017}.
Considering this, we selected to use a coastal setting for the relaxation experience in our experiment.
By making the sea breeze tangible using mid-air haptics, we aimed to create a holistic multisensory environment.
With permission by the content owners, we used a 360$^\circ$ stock video\footnote{www.atmosphaeres.com} showcasing a serene beach scene where the waves gently splash against the coast. 
An image of the scene is shown in Figure~\ref{fig:setup} a).
The footage also includes an audio track of the sea breeze.

In order to generate a congruent mid-air haptic sensation, we adopted a systematic iterative approach.
first we designed a sensation based on 
By taking into account the capabilities of the haptic device and principles grounded in physics (specifically, the consistent onshore movement of sea breezes during daytime, owing to the difference in air pressures arising from temperature disparities~\cite{Baines2003}), we designed an initial haptic stimulation pattern.
We used amplitude modulation to generate multiple haptic control points at a frequency of 200Hz.
To refine the haptic sensation, we subjected it to pilot testing with 5 users and iteratively improved it based on the feedback obtained. 
As a result of this preliminary study, we identified the following characteristics as closely approximating the users' perception of a sea breeze sensation (see Figure~\ref{fig:hand}):

\begin{enumerate}
	\item  The haptic sensation of the sea breeze encompasses 4 points of contact on the palm. The points are at a distance of 2.5cm from each other.
        \item The haptic sensation starts at an 8cm offset above the centre of the hand.
 	\item  The 4 points of contact are in a straight line and the line of haptic points moves from the tip of the fingers to the base of the palm in steps of 0.75cm, and with a delay of 0.1s. When the points reach the base of the palm, there is a 0.5s pause before they are generated again at the tip of the fingers.
	\item  The angle between the motion vector of the haptic stimulation (perpendicular to the line of haptic points) and the  y-axis of the palm, when oriented flat facing the mid-air haptic array, randomly varies every three iterations, adopting values from the range of [-45$^{\circ}$,45$^{\circ}$]. Figure~\ref{fig:hand} illustrates two possible arrangements of the control points, with the permissible range of angles denoted with grey lines. 
\end{enumerate}



\subsection{Measures and Procedure}
\begin{figure}[t]
  \centering
\includegraphics[width=\linewidth]{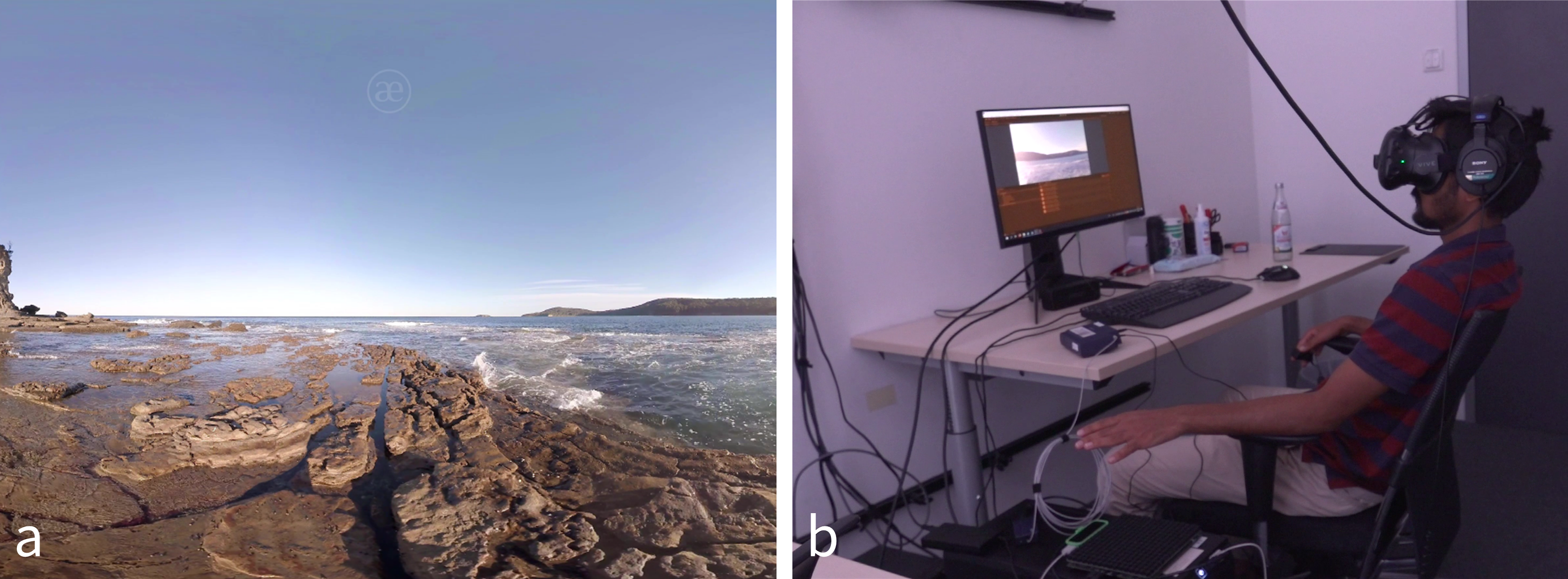}
  \caption{a) An image of the 360$^{\circ}$ video of a beach scene, which also included an audio track of the sea breeze. b) The image shows a participant experiencing the VR+Haptic condition. They wore a VR headset and extended one hand beyond the arm rest of the chair over the haptics board, which was positioned around 20 cm below the palm. The other hand was in a relaxed resting position.
  }
  \label{fig:setup}
\end{figure}

Upon arrival in the lab, the participant was provided with a document that contained basic information about the study.
They were encouraged to ask any questions they might have about the study or the technology used, and were reminded that they can quit the experiment at any time. 
Then they read and signed a consent form, and filled in a demographics questionnaire.
Next, the participant was seated on a comfortable chair with armrests, and they were instructed to sit still for 2 minutes to establish a baseline.
After this period passed, they were asked to fill in the pleasure and arousal dimension of the Self Assessment Manikin (SAM) questionnaire~\cite{Bradley1994}. 
Then the participant underwent a mild stress test. 
For this purpose we used the Short Sing a Song Stress Test (SSST)~\cite{djvandermee2020} that induces social-evaluative stress similar to that of Trier Social Stress Test~\cite{Kirschbaum1993} and the Sing a Song Stress Test~\cite{Brouwer2014}.
In particular, after reading some statements on the computer screen, the participant gets unexpectedly prompted to sing a song out loud for 60 seconds.
A slight variation of this stress task was used in~\cite{Kelling16}.
On completing the test, the participant filled in the SAM questionnaire again.
If the participant belonged to the control condition, they were asked to sit still for the next 5 minutes before the SAM questionnaire was administered for the third time.
Subsequently, a short interview about their experience was conducted.
Participants in the VR-only condition were instructed to put on the VR headset and the headphones, then they had 5 minutes to experience the VR scene. 
Participants in the VR+Haptics condition additionally received congruent mid-air haptic stimulation via the haptic device, placed approximately 20cm below their palm (see Figure~\ref{fig:setup} b)). 
After the 5 minutes passed, they were again asked to fill in the SAM questionnaire.
At the end of the experiment, we asked participants who were exposed to the VR scene 12 additional study-specific questions, to better understand how they experienced the relaxation intervention.
The questions were formulated along the example of the short form of the User engagement scale (UES)~\cite{Obrien18}, and similarly answered on a 5-point Likert scale (1 meaning \textit{not at all}, and 5 \textit{very likely/strongly agree}). 
The full list of questions is provided in Appendix~\ref{app_post-study-questionnaire}.

\begin{figure}[h]
  \centering
  \includegraphics[width=0.8\linewidth]{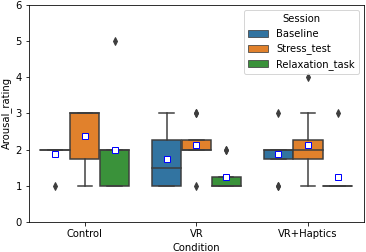}
  \caption{Boxplot of the Arousal dimension for the three conditions (Control, VR-only, and VR+Haptics).}
  \label{fig:sam_arousal}
\end{figure}

\begin{figure}[h]
  \centering
  \includegraphics[width=0.8\linewidth]{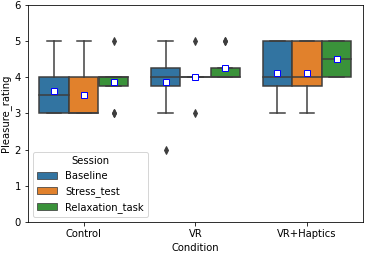}
  \caption{Boxplot of the Pleasure dimension for the three conditions (Control, VR-only, and VR+Haptics).}
  \label{fig:sam_pleasure}
\end{figure}

\section{Results}
\subsection{SAM}
The results of the SAM questionnaire for the arousal dimension are shown in Figure~\ref{fig:sam_arousal}. 
In the control condition, the initial mean arousal rating of 1.88 (SD 0.35) increased to 2.38 (SD 0.92) after the stress test, and decreased to 2.00 (SD 1.31) after the 5 minute rest. 
In the VR-only condition, the mean score of 1.75 (SD 0.89) after the baseline interval, increased to 2.13 (SD 0.64) after the stress test, and then decreased to 1.25 (SD 0.46) after the relaxation experience in VR.
Lastly, in the VR+Haptics condition, the mean score of 1.88 (SD 0.64) increased to 2.13 (SD 0.99) after the stress test, and then decreased to 1.25 (SD 0.71) after the relaxation experience.
The Kruskal-Wallis test indicated no significant difference in the arousal ratings between the different relaxation conditions ($\chi^2=4.22$, $df=2$,  $p=0.12$).

Figure~\ref{fig:sam_pleasure} shows the ratings in the pleasure dimension of the SAM questionnaire. 
In the control condition, the mean pleasure rating of 3.63 (SD 0.74) slightly decreases to 3.5 (SD 0.76) after the stress test, and increased to 3.88 (SD 0.64) after the rest period.
In the VR-only condition, the mean pleasure was 3.88 (SD 0.99) after the baseline. 
Then it slightly increased to 4.00 (0.53) after the stress test, and increased again to 4.25 (SD 0.46) after the VR relaxation experience. 
In the VR+Haptics condition, the mean pleasure rating of 4.13 (SD 0.83) remained the same after the stress test, and increased to 4.5 (SD 0.53) after the relaxation in VR with mid-air haptic feedback.
The Kruskal-Wallis test indicated a potential trend in the pleasure ratings between the conditions ($\chi^2=5.90$, $df=2$,  $p=0.05$), suggesting a potential pattern or tendency among the conditions. 
To confirm the significance of this trend, further investigation with a larger sample size is necessary.



\begin{figure*}[t]
  \centering
  \includegraphics[width=\linewidth]{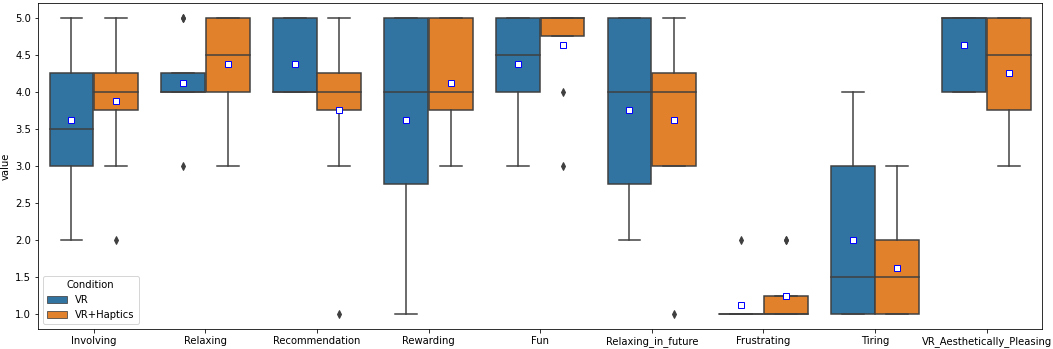}
  \caption{Boxplot of the responses to the questions 1 to 9 on the additional study-specific questionnaire for the VR-only and the VR+Haptics condition.}
  \label{fig:post_study_questions}
\end{figure*}

\subsection{Additional Questions}
\begin{figure}[h]
  \centering
  \includegraphics[width=0.8\linewidth]{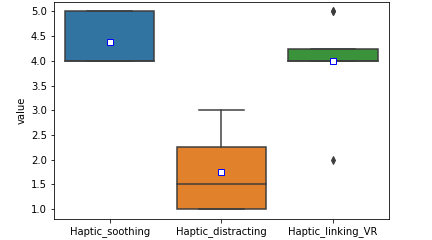}
  \caption{Boxplot of the responses to the questions 10 to 12 on the additional study-specific questionnaire for the VR+Haptics condition.}
  \label{fig:post_study_questions_haptics}
\end{figure}

The results of the additional study-specific questionnaire that we administered after the VR-only and the VR+Haptics conditions, to better understand how participants experienced the relaxation procedure is shown in Figure~\ref{fig:post_study_questions}.
For the VR-only condition, the participants rated the experience as involving (in terms of losing track of time) with a mean score of 3.63 (SD 1.06), relaxing with 4.13 (SD 0.64), would recommend it to others with 4.38 (0.52), rewarding with 3.63 (SD 1.51), fun with 4.38 (SD 0.74), frustrating with 1.13 (SD 0.35), and tiring with 2.00 (SD 1.20).
The VR+Haptics condition was rated as as involving with a mean score of 3.88 (SD 0.99), relaxing with 4.38 (SD 0.74), would recommend it to others with 3.75 (1.28), rewarding with 4.13 (SD 0.83), fun with 4.63 (SD 0.74), frustrating with 1.25 (SD 0.46), and tiring with 1.63 (SD 0.74).
Using the Kruskal-Wallis test, we did not find a significant difference between the conditions (p$>$0.05).
The last three questions of the questionnaire were related to the experience with the mid-air haptic feedback, hence they were only answered by the participants in the VR+Haptics condition. 
The results are presented in Figure~\ref{fig:post_study_questions_haptics}. 
The participants rated the haptic feedback as soothing with a mean score of 4.38 (SD 0.52), and distracting with 1.75 (SD 0.89). 
Lastly, participants rated the correspondence of the haptics with the VR scene with a mean score of 4.00 (SD 0.93).

\section{Discussion}
Our study focused on exploring the potential of a multimodal relaxation experience, combining a VR scene with congruent mid-air haptic feedback, in alleviating stress. 
We discuss our findings in relation to existing literature and implications for future stress management and relaxation applications.

We combined the immersive quality of VR with the potential benefits of mid-air haptic technology, hypothesising that this integration would lead to heightened relaxation and stress reduction.
While we were not able to confirm this hypothesis in this initial investigation, we did observe some interesting trends that warrant further investigation with a larger sample size.
Specifically, the results indicate a trend suggesting that the VR+Haptic condition might be associated with higher pleasure ratings compared to the other two conditions.
On average, participants rated the experience in the VR+Haptics condition as more involving, relaxing, and fun, while being less tiring, compared to the VR-only condition.
However, they hesitated about recommending the VR+Haptics experience to their friends and family, as they expressed some concerns regarding the price of the haptics hardware as well as operating it on their own in everyday life.
Similarly, the higher average frustration score for the VR+Haptics condition might be due to the unfamiliarity with the haptics board.
Although in the study, we observed that after some short instructions, the participants were able to correctly position their hand and optimally obtain the mid-air haptic stimulus.
Participants in the VR+Haptics condition found the mid-air haptic feedback to be soothing, and well-aligned with the VR scene, and most of them did not perceive it as distracting.

There are several study limitations that should be considered. 
Firstly, the relatively small sample size, consisting mainly of university students and employees, might limit the generalisability of our findings.
Furthermore, our study's focus on immediate effects and participants' immediate perceptions limits our ability to assess the sustained impact over time. 
Longitudinal studies are required to determine the real-world effectiveness of the relaxation experience in stress management over extended time periods. 
Secondly, individual differences and expectations, stemming from prior experiences and personal preferences, could have influenced participants' interaction with the interventions.
For instance, the Sing a Song Stress Test might not have been as effective with participants who have experience in performing arts.
This might partially explain the unchanged mean pleasure score in the VR+Haptics condition and the higher mean pleasure scores in the VR-only condition after the stress test, compared to the baseline scores.
Alternative stress-inducing tasks, such as the mental arithmetic stress test~\cite{Birkett2011}, could be considered for future studies.
Thirdly, the design of the mid-air haptic feedback pattern, while rooted in iterative preliminary testing, may not account for individual variability in sensitivity and perception.
For example, age-related differences in haptic sensitivity might exist.
This highlights the potential for customisation to optimise the relaxation experience.
Finally, augmenting the VR scene with additional (possibly interactive) elements, such as visual breathing guidance~\cite{Soyka2016} or a virtual breathing coach~\cite{Dar2022}, might further increase user engagement and immersion in the relaxation experience. 
Implementing interactive breath guidance using the mid-air haptic feedback offers yet another promising avenue for future exploration.


\section{Conclusion}

This study investigated for the first time the impact of using congruent mid-air haptic feedback in combination with a natural scene in VR, with the aim of stress reduction and relaxation.
The user study encompassed three different conditions: a control condition with no relaxation intervention, a VR-only, and a VR+Haptics relaxation experience.
Our findings did not reveal a significant difference between the conditions, however, we did observe a trend in the pleasure ratings that requires further investigation.
The feedback from the participants regarding the VR relaxation experience enhanced with mid-air haptics was favourable, as they found the experience to be immersive, pleasurable, fun, and relaxing.
These initial findings suggest that the integration of mid-air haptic feedback has the potential to enhance, rather than diminish, the virtual relaxation experience. 
In the future, the user study could be revisited with a larger and more diverse participant pool,  incorporating longitudinal assessments, enabling customisation of the haptic stimulation, and potentially pairing it with breathing exercises. 
Moreover, physiological data, such as heart rate and electrodermal activity, could be recorded as an additional measure of the participants' state.
Lastly, to accommodate individual preferences regarding relaxation, different scenes (e.g., mountains, lakes, etc.) with corresponding haptic sensations could be designed and compared.







\acknowledgments{
The authors would like to thank all participants in the user study.}

\bibliographystyle{abbrv-doi}
\bibliography{template}

\appendix

\section{Post-study Likert scale questionnaire}
\label{app_post-study-questionnaire}

\begin{tabular}{lcccccl}
\multicolumn{7}{l}{1. I was so involved in this experience that I lost track of time.}\\
strongly disagree & 1 & 2 & 3 & 4 & 5 & strongly agree\\
\multicolumn{7}{l}{}\\
\multicolumn{7}{l}{2. The experience was relaxing.}\\
strongly disagree & 1 & 2 & 3 & 4 & 5 & strongly agree\\
\multicolumn{7}{l}{}\\
\multicolumn{7}{l}{3. I would recommend the system to my family and friends.}\\
strongly disagree & 1 & 2 & 3 & 4 & 5 & strongly agree\\
\multicolumn{7}{l}{}\\
\multicolumn{7}{l}{4. This experience was rewarding.}\\
strongly disagree & 1 & 2 & 3 & 4 & 5 & strongly agree\\
\multicolumn{7}{l}{}\\
\multicolumn{7}{l}{5. This experience was fun.}\\
strongly disagree & 1 & 2 & 3 & 4 & 5 & strongly agree\\
\multicolumn{7}{l}{}\\
\multicolumn{7}{l}{6. I would want to use this system to relax in future.}\\
strongly disagree & 1 & 2 & 3 & 4 & 5 & strongly agree\\
\multicolumn{7}{l}{}\\
\multicolumn{7}{l}{7. I felt frustrated during the
experience.}\\
strongly disagree & 1 & 2 & 3 & 4 & 5 & strongly agree\\
\multicolumn{7}{l}{}\\
\multicolumn{7}{l}{8. I felt tired
after using the system.}\\
strongly disagree & 1 & 2 & 3 & 4 & 5 & strongly agree\\
\multicolumn{7}{l}{}\\
\multicolumn{7}{l}{9. The VR scene was
aesthetically pleasing.}\\
strongly disagree & 1 & 2 & 3 & 4 & 5 & strongly agree\\
\multicolumn{7}{l}{}\\
\multicolumn{7}{l}{10. I found the haptic
feedback soothing.}\\
strongly disagree & 1 & 2 & 3 & 4 & 5 & strongly agree\\
\multicolumn{7}{l}{}\\
\multicolumn{7}{l}{11. I found the haptic feedback
distracting.}\\
strongly disagree & 1 & 2 & 3 & 4 & 5 & strongly agree\\
\multicolumn{7}{l}{}\\
\multicolumn{7}{l}{12. The haptic sensation corresponds well with the VR scene.}\\
strongly disagree & 1 & 2 & 3 & 4 & 5 & strongly agree
\end{tabular}



\end{document}